# Temporal Analysis and Gender Bias in Computing

Thomas J. Misa
University of Minnesota
tmisa@umn.edu

ABSTRACT: Recent studies of gender bias in computing use large datasets involving automatic predictions of gender to analyze computing publications, conferences, and other key populations. Gender bias is partly defined by software-driven algorithmic analysis, but widely used gender prediction tools can result in unacknowledged gender bias when used for historical research. Many names change ascribed gender over decades: the "Leslie problem." Systematic analysis of the Social Security Administration dataset—each year, all given names, identified by ascribed gender and frequency of use—in 1900, 1925, 1950, 1975, and 2000 permits a rigorous assessment of the "Leslie problem." This article identifies 300 given names with measurable "gender shifts" across 1925–1975, spotlighting the 50 given names with the largest such shifts. This article demonstrates, quantitatively, there is net "female shift" that likely results in the overcounting of women (and undercounting of men) in earlier decades, just as computer science was professionalizing. Some aspects of the widely accepted 'making programming masculine' perspective may need revision.

Computing as a pervasive activity in the contemporary world faces two systemic problems, and historians of computing have significantly shaped public understanding of one of them. Presently, there is great interest in analyzing the power relations that shape society and culture through computing systems and networks, which the burgeoning literature on "algorithmic bias" offers to define and diagnose. Historians of computing have even more effectively shaped public understanding of gender bias in computing, an equally pernicious problem. It is a signal success when history scholars influence public debate extending from scholarly publications through mass media and documentary film. This article reassesses some of the evidence used to support this perspective.

The vivid tagline "making programming masculine" is associated with this dominant perspective in history of computing. As reviewed below in greater detail, scholars such as Jennifer Light, Nathan Ensmenger, and many others have suggested that gender bias in computing has been systemic, pervasive, and durable. The starting point was Light's suggestion that, originally, "programming was a woman's job." Subsequently, men's efforts to professionalize computing in the 1960s and 1970s, in Ensmenger's several publications, reportedly drove women out of computing and consequently resulted in a professional field consistently and continually dominated by men. It seems a straightforward path from there to the egregious examples of sexism in the contemporary computer industry. Such a perspective posits gender bias as an enduring "structure" in professional male-dominated computing, fully in line with many women's studies and gender scholars who also examine the roots of contemporary gender-biased practices and norms.

The current enthusiasm for digital humanities encourages inquiry across immense textual, bibliometric, image, or other datasets, or corpora.[1] The significant gender bias in natural-language corpora on which machine learning programs operate led, infamously, to the meme "Man is to computer programmer as woman is to homemaker."[2] Correcting or mitigating such word-embedding bias is a significant problem now recognized in computer science.[3] A similar caution concerns the rise of digital humanities and digital history.[4] The "objects" of digital humanities inquiry—words, phrases, names, patents—sometimes change character across the timeframe under investigation.[5] Computers, unaided by human insight, do an inadequate job at assessing context, discerning changes across time, and grounding robust interpretive humanities results. In sum, there is likely to be significant advantage in keeping humans with significant contextual knowledge "in the loop" when using software tools in historical research.

[1] For analysis of the Computational Chemistry [email] List corpora, see Alexandre Hocquet and Frédéric Wieber, "'Only the Initiates Will Have the Secrets Revealed': Computational Chemists and the Openness of Scientific Software," *IEEE Annals of the History of Computing* 39 no. 4 (2017): 40-58 at doi: 10.1109/MAHC.2018.1221048.

[2] Tolga Bolukbasi, Kai-Wei Chang, James Zou, Venkatesh Saligrama, and Adam Kalai, "Man is to Computer Programmer as Woman is to Homemaker? Debiasing Word Embeddings," (2016) available at arxiv.org/abs/1607.06520v1 and elsewhere.

[3] Marzieh Babaeianjelodar, Stephen Lorenz, Josh Gordon, Jeanna Matthews, and Evan Freitag, "Quantifying Gender Bias in Different Corpora," In Companion Proceedings of the Web Conference 2020 (WWW '20). Association for Computing Machinery, New York, NY, 2020, 752–759 at doi.org/10.1145/3366424.3383559; Joshua Gordon, Marzieh Babaeianjelodar, and Jeanna Matthews, "Studying Political Bias via Word Embeddings," Companion Proceedings of the Web Conference 2020 (WWW '20). Association for Computing Machinery, New York, NY, 2020, 760–764 at doi.org/10.1145/3366424.3383560; Ninareh Mehrabi, Thamme Gowda, Fred Morstatter, Nanyun Peng, and Aram Galstyan, "Man is to Person as Woman is to Location: Measuring Gender Bias in Named Entity Recognition," In Proceedings of the 31st ACM Conference on Hypertext and Social Media (HT '20). Association for Computing Machinery, New York, NY, 2020, at doi.org/10.1145/3372923.3404804

[4] For forecasts of the digital future of history see Thomas Bartscherer and Roderick Coover, eds., *Switching Codes: Thinking Through Digital Technology in the Humanities and the Arts* (Chicago: University of Chicago Press, 2011), which posits "technologies that are so radically transforming [scholars'] fields" (quote p. 2); William J. Turkel, Shezan Muhammedi, and Mary Beth Start, "Grounding Digital History in the History of Computing," *IEEE Annals of the History of Computing* 36, no. 2 (2014): 72-75 at doi: 10.1109/MAHC.2014.21. Compare Marc Weber, "Self-Fulfilling History: How Narrative Shapes Preservation of the Online World," *Information & Culture* 51 no. 1 (2016): 54-80 at doi:10.1353/lac.2016.0003.

This article begins with a brief overview of recent historical work in the dynamics of gender bias in computing. It next examines several commonly used digital-humanities-type tools (readily available through application programming interfaces or APIs) for automatically inferring gender from given or first names. It then analyzes a sample of 13 million named individuals based on the comprehensive Social Security Administration dataset of all US given first names tabulated by name-frequency (year by year) and assigned gender at birth. "Gender switching" over decades is surprisingly common. What is the impact? The present analysis demonstrates that the "net" female-gender shift results in female over-counting (e.g. an historical Leslie likely to be male mis-identified as a female) is far larger than a countervailing male-gender shift. Focused analysis of 133 persons named Leslie publishing in computer science (1970–2020), identified in the ACM Digital Library, illustrates this temporal shift. The conclusion suggests that both methods discussed in this article—human-centered methods that are temporally and contextually grounded as well as large-scale computer-driven gender predictions—are likely to be valuable in historical research on computing (and other topics) when they are combined.

Gender scholars sometimes offer a "cultural" argument in which statistics or numbers in a population are simply of little account. They note quite correctly that women can be numerically gaining access to a professional field, and so their numbers might be rising, while still being excluded from positions of power and influence. Numbers, alone, are not definitive in a cultural argument. The scholars advocating the "making programming masculine" perspective, however, have asserted (thinly sourced) sociological facts such as Ensmenger's conjecture that women were "30 percent" or more of early computer programmers prior to male-driven exclusionary professionalization. If computing has, for six decades, been saturated with sexism it seems difficult to imagine contemporary reform movements strong enough to alter such an enduring structure. But what if computing during these decades was not *continually* saturated with sexism? There is significant public purpose in evaluating and, if needed, correcting the dominant public view of gender bias in computing. I favor an alternative perspective that gender bias was not inherently baked into professional computing but rather a contingent feature that emerged in the mid-1980s—well after professionalization in the 1960s and 1970s. My approach recommends careful attention to populations and datasets that have been little acknowledged in history of computing (three recent publications can be added to this article after peer review).

[5] For case studies, see Douglas O'Reagan and Lee Fleming, "The FinFET Breakthrough and Networks of Innovation in the Semiconductor Industry, 1980–2005: Applying Digital Tools to the History of Technology," *Technology and Culture* 59 no. 2 (2018): 251-288 at doi:10.1353/tech.2018.0029; and Shannon Jackson, "The City from Thirty Thousand Feet: Embodiment, Creativity, and the Use of Geographic Information Systems as Urban Planning Tools," *Technology and Culture* 49 no. 2 (2008): 325-346 at doi: 10.1353/tech.0.0039; Lisa Onaga and Hanna Rose Shell, "Digital Histories of Disasters: History of Technology through Social Media," *Technology and Culture* 57 no. 1 (2016): 225-230. doi: 10.1353/tech.2016.0020.

## Gender bias in computing

Historians of computing, beginning with Jennifer Light, helped identify and define historical perspectives on gender bias in computing. In 1999 Light published "When Computers Were Women" in *Technology and Culture*, identifying an idiom of sex-typing that was common during and following World War II: her suggestion was that "designing [computer] hardware was a man's job; programming was a woman's job." Ground zero was assessing the pioneering women of ENIAC, who played critical—and belatedly recognized—roles in getting ENIAC to do useful work, or what now is appreciated as programming the early machine. Light suggested that "the job of programmer, perceived in recent years as masculine work, originated as feminized clerical labor."[6]

The first decade of computer programming had numerous prominent women, including Frances Holberton, Jean Jennings Bartik, Ida Rhodes, and the incomparable Grace Hopper, while Mina Rees, Florence Koons, and many other women also contributed to early computing, as numerous studies in *Annals* and elsewhere have shown. Clearly, women gained striking prominence; some historians even conjectured they were as much as half of the emerging field. "The exact percentage of female programmers is difficult to pin down with any accuracy," wrote Nathan Ensmenger, "but … reliable contemporary observers suggest that it was [close] to 30 percent." Elsewhere, with anecdotal evidence, he suggested women were possibly *50* percent of computer programmers in the years before male-dominated professionalization resulted in "making programming masculine." Subsequently, Ensmenger suggested "the masculinization of computer programming" shaped and sharpened computing's pervasive masculine culture. "Over the course of the 1960s and 1970s, male computer experts were able to successfully transform the 'routine and mechanical' (and therefore feminized) activity of computer programming into a highly valued, well-paying, and professionally respectable discipline," he suggested. "They did so by constructing for themselves a distinctively masculine identity."[7]

---

[6] Jennifer S. Light, "When Computers Were Women," *Technology and Culture* 40 no. 3 (1999): 455-483, quotes on 455 (job of programmer) and 469 (man's job) at www.jstor.org/stable/25147356. See W. Barkley Fritz, "The Women of ENIAC," *IEEE Annals of the History of Computing* 18 no. 3 (1996): 13-28 at doi.org/10.1109/85.511940

Light and Ensmenger's findings have appeared widely in academic and popular media, including National Public Radio, the *Wall Street Journal*, and Robin Hauser Reynolds' acclaimed documentary "Code: Debugging the Gender Gap" (2015), which prominently voiced Ensmenger's suggestions, including the claims of 30 to 50 percent women.[8] Liza Mundy in the *Atlantic* magazine summarizes the now widely accepted view: "after World War II, software programming was considered rote and unglamorous, somewhat secretarial—and therefore suitable for women. The glittering future, it was thought, lay in hardware. But once software revealed its potential—and profitability—the guys flooded in and coding became a male realm."[9]

Yet recent studies based on systematic and/or longitudinal data do not confirm the "making programming masculine" conjecture. Quite the reverse. Whereas the conjecture would hypothesize that women must become a smaller proportion of the computing community (by whatever measure) during the years of male-driven exclusionary professionalization, roughly, the 1950s through 1970s, studies based on systematic data point in the opposite direction. Based on 52 interviews with US and UK women computing pioneers, Janet Abbate writes that "women . . . held positions of responsibility and influence in the early computer industry . . . and were employed in numbers that, while a small minority of the total, compared favorably with women's representation" in other areas of science and engineering at the time." Women earned an increasing share of computer-science bachelor's degrees from the mid-1960s through the mid-1980s; and, likewise, women constituted an increasing share of the highly skilled white-collar computing workforce, according to widely available data from the National Science

---

[7] Nathan Ensmenger, "Making Programming Masculine," in Thomas J. Misa, ed., *Gender Codes: Why Women Are Leaving Computing* (John Wiley, 2010), 115-141, quote p. 116. For the claim of 50 percent, see Ensmenger at https://web.archive.org/web/20180105182302/homes.soic.indiana.edu/nensmeng/files/ensmenger-gender.pdf on p. 2; Nathan Ensmenger, "'Beards, Sandals, and Other Signs of Rugged Individualism': Masculine Culture within the Computing Professions," *Osiris* 30 (2015): 38-65 at doi.org/10.1086/682955 p. 38 (quoted in text above).

[8] See (e.g.) Laura Sydell, "The Forgotten Female Programmers Who Created Modern Tech," NPR Morning Edition (6 October 2014) at www.npr.org/sections/alltechconsidered/2014/10/06/345799830/the-forgotten-female-programmers-who-created-modern-tech; Christopher Mims, "The First Women in Tech Didn't Leave—Men Pushed Them Out," *Wall Street Journal* (10 December 2017) at www.wsj.com/articles/the-first-women-in-tech-didnt-leavemen-pushed-them-out-1512907200; Stephen Cass, "A Review of *Code: Debugging the Gender Gap*," *IEEE Spectrum* (19 June 2015) at spectrum.ieee.org/geek-life/reviews/a-review-of-code-debugging-the-gender-gap

[9] Mundy, L. "Why is Silicon Valley so Awful to Women?" *The Atlantic* (April 2017): 60-73 at web.archive.org/web/20201217090534/www.theatlantic.com/magazine/archive/2017/04/why-is-silicon-valley-so-awful-to-women/517788/

Foundation and Census Bureau.[10] A 2011 study by Cohoon, Nigai, and Kaye found that women researchers authored a growing share of computer-science publications, increasing from 7% of authors in 1967 to 27% in 2009. This increase roughly parallels the increasing number of women gaining computer-science Ph.D. degrees.[11] And a massive study by Lucy Wang and colleagues found that women similarly increased their proportion of computer-science authorship, from around 12% in the 1960s through to a peak of just under 30% in recent years (2015-2020).[12] During roughly the 1960s through the mid-1980s, the view from multiple large-scale, systematic datasets affirms that women were flooding into computer science and the wider computing profession. Historians of computing have not adequately addressed this disconnect between systematic and anecdotal data.

## Inferring gender through computer analysis

The six women of ENIAC were initially a significant number for historians of computing to consider, but today the datasets used in gender analysis of computing are much larger. One study assessed 1,285 letters to an early personal-computer magazine.[13] A recent bibliometric analysis is based on nearly 18,000 names.[14] The author published an longitudinal analysis based on 50,000 individuals. A 2011 study of computer-science conference papers drew on a dataset of 350,000.[15] In 2021 an assessment of historical trends in computer science publishing relied on a dataset of 11.8 million computer-science publications.[16] An emerging cottage industry compares and assesses various methods, datasets, and algorithms used in these software tools for automatic gender identification.[17]

---

[10] Data discussed in Caroline Clarke Hayes, "Computer Science: The Incredible Shrinking Woman," in Thomas J. Misa, ed., *Gender Codes: Why Women are Leaving Computing* (Hoboken, NJ: Wiley, 2010), pp. 25-49 at https://doi.org/10.1002/9780470619926.ch2

[11] J. McGrath Cohoon, Sergey Nigai, and Joseph 'Jofish' Kaye, "Gender and Computing Conference Papers," *Communications of the ACM* 54 no. 8 (2011): 72–80. DOI: https://doi.org/10.1145/1978542.1978561

[12] Lucy Lu Wang, Gabriel Stanovsky, Luca Weihs, and Oren Etzioni, "Gender Trends in Computer Science Authorship," *Communications of the ACM* 64 no. 3 (2021): 78–84. DOI: https://doi.org/10.1145/3430803

[13] Laine Nooney, Kevin Driscoll, and Kera Allen, "From Programming to Products: *Softalk* Magazine and the Rise of the Personal Computer User," *Information & Culture* 55 no. 2 (2020): 105-129 at muse.jhu.edu/article/757558.

[14] Sandra Mattauch, Katja Lohmann, Frank Hannig, Daniel Lohmann, and Jürgen Teich, "A Bibliometric Approach for Detecting the Gender Gap in Computer Science," *Communications of the ACM* 63 no. 5 (2020): 74–80 at https://doi.org/10.1145/3376901

[15] J. McGrath Cohoon, Sergey Nigai, and Joseph 'Jofish' Kaye, "Gender and Computing Conference Papers," *Communications of the ACM* 54 no. 8 (2011): 72–80 at https://doi.org/10.1145/1978542.1978561

[16] Lucy Lu Wang, Gabriel Stanovsky, Luca Weihs, and Oren Etzioni, "Gender Trends in Computer Science Authorship," *Communications of the ACM* 64 no. 3 (2021): 78–84 at https://doi.org/10.1145/3430803

Several gender analysis tools are readily accessible. One of the earliest was Genderyzer used to assess authorship in computer science publications.[18] Others similar tools are Gender API, NameAPI, NamSor, Genderize.io, Gender-Guesser, Gender REST API,[19] and genderizeR[20] which advocates "using informal, crowd-sourced and not widely recognized data sources" as a means to predict gender from given names. Some tools such as genderizeR focus on a restricted set of the top-1000 SSA names (rather than the significantly larger complete SSA dataset) with an aim of "predicting" gender, so that it considers only names with a female-or-male gender probability >0.95. This analytical model presumes that gender is binary in contrast with the probabilistic conception of gender advocated in this article. Several publications consider historical assessment in the context of bibliometric analysis, sometimes with an implied gender-binary model.[21]

---

[17] Fariba Karimi, Claudia Wagner, Florian Lemmerich, Mohsen Jadidi, and Markus Strohmaier, "Inferring Gender from Names on the Web: A Comparative Evaluation of Gender Detection Methods," In Proceedings of the 25th International Conference Companion on World Wide Web (WWW '16 Companion). International World Wide Web Conferences Steering Committee, Republic and Canton of Geneva, CHE, 53–54, at doi.org/10.1145/2872518.2889385 ; Hua Zhao and Fairouz Kamareddine, "Advance Gender Prediction Tool of First Names and its Use in Analysing Gender Disparity in Computer Science in the UK, Malaysia and China," 2017 International Conference on Computational Science and Computational Intelligence (CSCI), Las Vegas, NV, USA, 2017, pp. 222-227, doi: 10.1109/CSCI.2017.35 ; Lucía Santamaría and Helena Mihaljević, "Comparison and Benchmark of Name-to-Gender Inference Services," *PeerJ Computer Science* (July 16, 2008) at http://dx.doi.org/10.7717/peerj-cs.156 ; Foad Hamidi, Morgan Klaus Scheuerman, and Stacy M. Branham, "Gender Recognition or Gender Reductionism? The Social Implications of Embedded Gender Recognition Systems," In Proceedings of the 2018 CHI Conference on Human Factors in Computing Systems (CHI '18). Association for Computing Machinery, New York, NY, USA, Paper 8, 1–13. DOI: https://doi.org/10.1145/3173574.3173582 ; Stefan Krüger and Ben Hermann, "Can an Online Service Predict Gender? On the State-of-the-Art in Gender Identification From Texts," In Proceedings of the 2nd International Workshop on Gender Equality in Software Engineering (GE '19). IEEE Press, 13–16. DOI: https://doi.org/10.1109/GE.2019.00012

[18] J. McGrath Cohoon, Sergey Nigai, and Joseph 'Jofish' Kaye, "Gender and Computing Conference Papers," *Communications of the ACM* 54 no. 8 (2011): 72–80. DOI: https://doi.org/10.1145/1978542.1978561. Genderyzer now at jofish.com/cgi-bin/genderyze.py

[19] https://www.programmableweb.com/api/gender-rest-api

[20] Kamil Wais, "Gender Prediction Methods Based on First Names with genderizeR," *The R Journal* 8 no. 1 (August 2016) at https://journal.r-project.org/archive/2016-1/wais.pdf

[21] L. Mullen, "Predict Gender From Names Using Historical Data" 2014 at https://github.com/ropensci/gender; Sandra Mattauch, Katja Lohmann, Frank Hannig, Daniel Lohmann, and Jürgen Teich, "A Bibliometric Approach for Detecting the Gender Gap in Computer Science," *Communications of the ACM* 63 no. 5 (2020): 74-80 at https://doi.org/10.1145/3376901

Two of most widely used such tools are the highly regarded Gender-API and NamSor API.[22] Many published studies have used one or the other as a principal means to predict gender.[23] Based in France, NamSor has a database of 5.5 million names and was designed to predict gender, measure cultural origins, and assess the economic flows of globalization; it boasts worldwide coverage including "all languages, alphabets, countries, regions" while using diverse language scripts.[24] Gender-API was created in 2014 by a German social-media programmer and its current database has 6 million names from 189 countries. It is a powerful tool, advertised for optimizing online registration forms; and gender researchers avidly use it. "Gender API outputs the predicted binary gender (female or male)," along with the confidence of that prediction and the number of samples on which it is based, notes one study (Wang et al.). While admirably suited to its intended present-day purpose, Gender-API like NamSor has significant weakness when used as a historical tool to assess gender across time.

As table 1 shows, Gender-API confidently but incorrectly predicts ($p > 0.9$) that Sydney, Allison, Leslie, Courtney, and Haley are female names (among the ten top-ranked "gender switching" names listed in appendix 2); but in the Social Security Administration's comprehensive historical dataset (year-of-birth = 1925) only *Jean* can be so confidently identified as female. Historically, most of these top-10 "gender switching" names were substantially or even overwhelmingly male ($0.08 < p(F) < 0.20$), where p(F) is female-gender probability. Also shown in table 1, an historical analysis relying on NamSor would similarly and substantially over-count as female the historically male-inclining names Sydney, Allison, Leslie, Shelby, Courtney, Haley, Bailey, and Kelly (by various amounts as the differences in p(F) indicate). Finally, Genderize.io predicts most of the same names to be female, incorrectly, with lower seeming confidence; it pegs *Jean* as a male name. The predicted p(F) figures — across these three software tools — when examined name-by-name are almost laughably out of agreement.

The four p(F) columns in Table 1 demonstrate significant divergence between gender predictions from Gender-API, Nam-Sor, and Genderize.io versus the SSA's actual historical gender probabilities. It follows that any study using any of these gender-predicting software tools for *historical* research—including Wang et al.'s massive study of 13 million computer science publications and many others—will likely yield questionable historical results.[25]

[22] "Gender API is in general the best performer in our benchmarks, followed by NamSor," notes Lucía Santamaría and Helena Mihaljević, "Comparison and Benchmark of Name-to-Gender Inference Services," *PeerJ Computer Science* (July 16, 2008) at http://dx.doi.org/10.7717/peerj-cs.156, p. 2.

[23] See (among numerous titles) Muneera Bano and Didar Zowghi, "Gender Disparity in the Governance of Software Engineering Conferences," In Proceedings of the 2nd International Workshop on Gender Equality in Software Engineering (GE '19). IEEE Press, 21-24 at https://doi.org/10.1109/GE.2019.00016 ; Abdulhakim Qahtan, Nan Tang, Mourad Ouzzani, Yang Cao, and Michael Stonebraker, "Pattern Functional Dependencies for Data Cleaning," *Proc. VLDB Endow.* 13 no. 5 (January 2020), 684-697 at https://doi.org/10.14778/3377369.3377377; Mattauch et al., "A Bibliometric Approach for Detecting the Gender Gap in Computer Science" (op cit).

[24] See https://www.namsor.com/

[25] Wang et al.'s study actually deals also with 18 other scholarly fields, in addition to computer science, but, since it uses the same questionable gender-analysis scheme for all fields, the results show similar odd patterns (Figure 4 on p. 83)—across such gender-diverse fields as art, biology, business, chemistry, economics, engineering, geology, history, materials science, mathematics, philosophy, and sociology: with one exception (medicine) all fields decline in proportion of female authors from 1940 to the late 1960s, then, again, many fields anomalously "bump upward" in the single year 1970; and then all fields rise in proportion of female authorship through to 2020. One possibility for such results that the underlying gender-analysis *method*—i.e. use of Gender-API—may be the root cause for these odd findings. Wang et al. are analyzing 'names' rather than actual individuals.

Table 1: Gender-prediction API results (vs. actual SSA data)

| Name | Gender-API | p(F) from Gender-API | NamSor API | p(F) from NamSor API | Genderize.io | p(F) from Genderize.io | Actual p(F) from SSA yob=1925 |
|---|---|---|---|---|---|---|---|
| Sydney | F | 0.95 | F | 0.87 | M | 0.41 | 0.1623 |
| Jean | F | 0.75 | M | 0.37 | M | 0.05 | 0.9745 |
| Allison | F | 0.99 | F | 0.93 | F | 0.93 | 0.2093 |
| Leslie | F | 0.92 | F | 0.87 | F | 0.77 | 0.0839 |
| Shelby | F | 0.91 | F | 0.98 | F | 0.68 | 0.0657 |
| Courtney | F | 0.94 | F | 0.98 | F | 0.79 | 0.2063 |
| Willie | M | 0.05 | M | 0.41 | M | 0.12 | 0.3565 |
| Haley | F | 0.98 | F | 0.97 | F | 0.92 | 0.5833 |
| Bailey | F | 0.82 | F | 0.54 | M | 0.41 | 0.1667 |
| Kelly | F | 0.87 | F | 0.99 | F | 0.87 | 0.1168 |

### Gender-switching names

Gertrude Stein famously quipped "Rose is a rose is a rose is a rose," originally with a particular Rose in mind; but in fact many adults in the early years of digital computing (here, e.g., year-of-birth = 1920) were given *male* names that today would likely be understood as soundly female ones. These include Florence, Thelma, Virginia, Marie, Opal, Camille, Marian, Isabel, Edith, and Alice. Mary was preponderantly a female name as was Rose, even though 24 boys were named Rose in 1920. In 1920 an additional 24 *boys* each were named Evelyn, Holly, Irene, and Shellie (today female names); the same number of boys were named Freddy, Kevin, Hercules, and Melvyn (today male names). By comparison, Willie was the most-common given girl's name in 1920 that might today be identified as male. And a "Boy Named Sue" (Johnny Cash's #1 hit on the country music charts in 1969) was not unknown, with 7 such boys in 1917 alone.[26]

[26] "The Surprising Story Behind Johnny Cash's 'A Boy Named Sue'," Wide Open Country (2018) at https://www.wideopencountry.com/the-story-behind-johnny-cashs-most-famous-story-song-a-boy-named-sue/

There may be some means of analyzing the Social Security dataset to gain insight about non-binary gender identities,[27] but in this article I am interested in gaining systematic insight into changes in the numbers of women and men active in computing during the 20th century. For this I analyzed all individual names tabulated in the five quarter-century marks stretching from 1900 to 2000, based on SSA's listed year-of-birth (N = 13 million). My goal was to develop a quantitative measure to assess the underlying data used in gender-prediction tools (see also author's forthcoming publication). To anticipate, three widely used software tools will *overstate* the number of women in the early years of digital computing (roughly 1940s through 1960s) owing to the tools' assumption that gender distributions that are valid today can be applied to past decades; consequently, use of these tools generates unacknowledged gender bias in historical investigations.

These tools, as mentioned above, use a present-day snapshot of gender probabilities to "infer" or "predict" gender from given first names: they inadequately account for temporal shifts in the gender probabilities of given names. Across these decades *names change* their ascribed gender, sometimes dramatically; and, moreover, there are more names, and a far greater number of individuals, that experience a female-shift in gender than ones that do a countervailing male-shift. Use of these gender-analysis tools will over-estimate the number of women active in the early decades of computing and other scholarly fields. Recall that the "making programming masculine" conjecture posits substantially high numbers for women in early computing; otherwise, there are too few women to be excluded during male-driven professionalization. Accurate assessment is one of the evidential pivots for the two rival perspectives on gender bias in computing.

My systematic analysis of the SSA data proceeded as follows. For each of the five selected years (1900, 1925, 1950, 1975, and 2000) I extracted all name–gender–number triplets from the complete online SSA dataset, which begins in 1880. For instance in 2000, Abigail was used to name 13,088 girls and an additional 16 boys. The probability that an Abigail born that year was female, based on their given name, can be computed as 0.9987 (number of female Abigails ÷ total number of Abigails). Conversely, Ethan was used to name 21 girls and 15,223 boys for a female-gender probability, or p(F), of 0.0013. Common names in 2000 with female-gender probabilities close to 0.5 (gender ambiguous) include Aaryn, Amel, An, Arlen, Armani, Ashtin, and Avery (to take the A's alone with n>20). The SSA dataset counts all uses, identifying Female and/or Male for each name; names used fewer than five times in any given year are not listed.

---

[27] Diego Couto de Las Casas, Gabriel Magno, Evandro Cunha, Marcos André Gonçalves, César Cambraia, and Virgilio Almeida, "Noticing the Other Gender on Google+," In *Proceedings of the 2014 ACM conference on Web science* (*WebSci '14*). Association for Computing Machinery, New York, NY, USA, 156-160 at https://doi.org/10.1145/2615569.2615692; Katta Spiel, Os Keyes, and Pınar Barlas, "Patching Gender: Non-binary Utopias in HCI," In Extended Abstracts of the 2019 CHI Conference on Human Factors in Computing Systems (CHI EA '19). Association for Computing Machinery, New York, NY, USA, Paper alt05, 1–11 at https://doi.org/10.1145/3290607.3310425 ; and Mar Hicks, "Hacking the Cis-tem," *IEEE Annals of the History of Computing* 41 no. 1 (2019): 20-33 at doi: 10.1109/MAHC.2019.2897667.

One modest form of gender ambiguity in the SSA dataset is common. For the years 1925, 1950, and 1975, most newborn children (83-86%) were given names that were used for either gender, though not used in equal numbers. In 1900, just 55% of children were given (let's say) such gender-ambiguous names; while in 2000 69% of children were given such names. These names are not 'ambiguous' in the sense of today's non-gender-binary culture; more precisely, the gender assignment of these names is probabilistic. Some names were close to gender binary: in 1975 Abigail was used to name 614 girls (and fewer than 5 boys); Alfonso, 433 boys (and fewer than 5 girls).

For my analysis, I computed the probability that each name was given to a female child for each of the five sample SSA years. These female-gender probabilities, or p(F), are charted 1925–2000 (see figure 1) for the 24 top-most gender-shifting names weighted by the names' frequency of use. Unmistakably, names change gender.

Figure 1: Gender p(F) for top 24 names* [weighted by N] 1900-2000

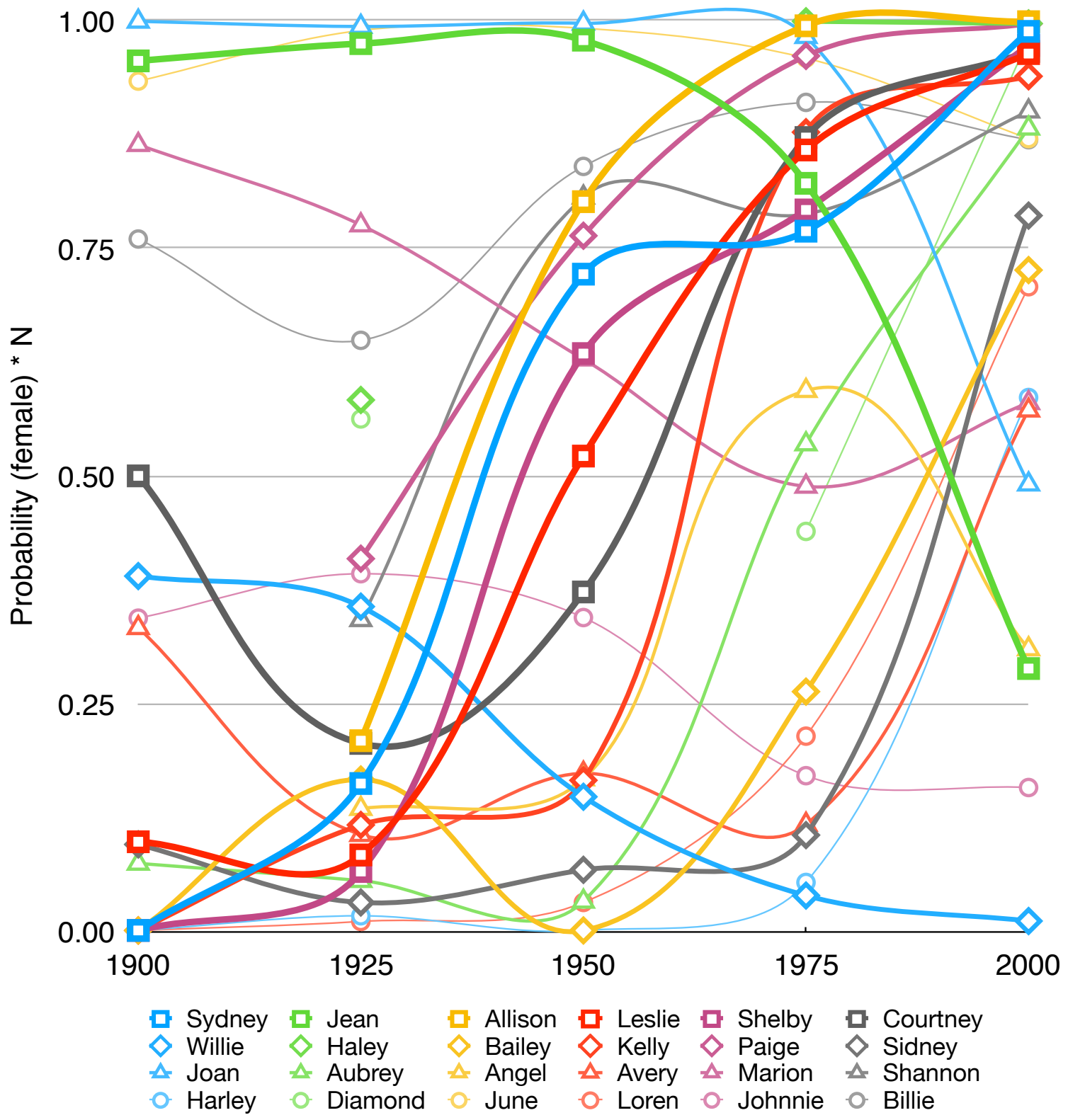

NB: "top 24" names with largest "gender switch" ∂p(F) from 1925 to 1975, weighted by frequency of names' use. Names identified in Appendix 2.

Multiple rounds of spreadsheet sorting, matching, and filtering allowed me to measure, name by name, the gender-probability *changes* between 1925 and 2000 for all names in the SSA dataset. Additional sorting pinpointed the top 50 *names* that underwent the largest gender shifts 1925–2000 (see Appendix 1). I also identified the top 50 gender-shifting names weighted by *frequency* of use (see Appendix 2). So while the given names Shelby, Leslie, Aubrey, and Sydney showed the greatest gender shift, by name alone, Sydney, Jean, Allison, and Leslie showed the great shift when weighted by frequency of use. The list of top gender-switching names weighted by frequency is the most consequential for diagnosing possible bias in existing gender analysis tools.

Then, I computed the gender-shift of each name during 1925–2000 by subtracting the gender probabilities p(F) for each name in these two years; since some of the shifts were small, for convenience I multiplied delta-p(F) by 100 to create legible numbers (see Appendix 1). Of the top 20 names, only Robbie, Jean, and Lavon experienced a 'negative' or male shift in p(F): these names became more associated with males from 1925 to 2000. The entire top 10 and the vast majority of the top-50 names experienced a 'positive' or female shift in p(F): these names became more associated with females from 1925 to 2000. The *median* across both positive and negative shifts is +55.36, indicating that 'positive' female-shifts outnumber 'negative' male-shifts. The preponderance of 'positive' shift in p(F) indicates that males—including computer-science historical figures born around and after 1925—are unwittingly mis-identified as female today. This difference can be used to define a quantitative measure to assess the gender analysis software tools over time (in the following section).

The name Shelby experienced significant gender shift, but historically consequential are the gender shifts in more-frequently used names such as Sydney, Jean, and Leslie. Once again, while there are some frequency-weighted names experiencing 'negative' shift in p(F) such that Jean, Willie, Joan, Marion and others became less common female names and more common male names, the preponderance of frequency-weighted names experienced 'positive' female gender shifts, such as Sydney, Allison, Leslie, Shelby, and many others. In the top-50, as well as the complete population of just over 300 names, more names experienced positive female-gender shift. The overall shift in frequency-weighted p(F) was strongly positive in the top-50, with an average across both positive and negative 'shift' numbers of +704.

## Consequences for API-based gender?

Absent reanalysis of the data, it is not possible to correct the figures in previously published studies, cited above. Several observations suggest there is a significant, systemic, and unreported bias in gender predictions using these tools. As noted above, names assigned to male and female babies were surprisingly common during 1925–1975 (83 to 86% of all names), during the very years of professionalization of computing and the possible onset of gender bias in computing. My analysis demonstrates that gender shift (positive or negative) occurred in more than 300 names—the top-50 are just that. Names in the top-50 most commonly made substantial male-to-female "switches": for instance, Shelby changed from a small 0.0657 probability of being female in 1925 to a large 0.9725 probability in 2000. To repeat, a historical person named Shelby in the early decades of professional computing was most likely a man—but today would be mis-identified as a woman by gender prediction software tools.

One such person is Shelby Brumelle, a business and information systems scholar at the University of British Columbia. Brumelle gained two publications coauthored with Allan J. Humphrey in 1966 for "An Algorithm for Retrieving Indexed Documents and its Application" initially presented at ACM's national conference and then published in *Communications of the ACM*; the computer-science index DBLP lists 13 additional publications.[28] 'Shelby' would be confidently identified by Gender-API and NamSor as female (0.91 < p(F) < 0.97), except that this Shelby was male.[29] He is not alone. In computing, the given name Leslie is naturally associated with ACM Turing Award laureates Leslie Valiant (2010), the influential computer-science and machine-learning theorist, and Leslie Lamport (2013), inventor of LaTeX and theorist of complex distributed systems. Like Brumell, Valiant and Lamport are unmistakably male.[30]

Besides Valiant and Lamport, the ACM Digital Library identifies 131 *additional* Leslie's as published authors (1970–2020). What can we say about these computer-science Leslie's? For decades, historians have identified their not-so-famous human subjects by piecing together biographical data from publications, testimony of colleagues, career summaries, correspondence, newspapers, magazines, photographs, public events, obituaries, spade-work in indexes or biographical dictionaries, and sometimes plain luck. Today we have also Google searches, Linked-In's career summaries (and peer recommendations), social media, and Internet Archive as well as personal and institutional websites. Computer science authors are not shy and retiring; personal data from themselves, students, and colleagues are everywhere.

---

[28] Allan J. Humphrey and Shelby L. Brumelle, "An Algorithm for Retrieving Indexed Documents and its Application," In Proceedings of the 1966 21st national conference (ACM '66). Association for Computing Machinery, New York, NY, USA, 499–504 at doi.org/10.1145/800256.810731 ; Allan J. Humphery and Shelby L. Brumelle, "An Algorithm for Retrieving Indexed Documents and its Application," *Communications of the ACM* 9 no. 7 (1966): 483 at doi.org/10.1145/365719.366481; compare https://dblp.org/pid/81/6367.html

[29] See Brumelle's biography at www.jstor.org/stable/171662 *or* https://students.ubc.ca/enrolment/finances/award-search/vancouver/sauder-school-business/general/6388 *or* open.library.ubc.ca/collections/arphotos/items/1.0214062

[30] See https://amturing.acm.org/award_winners/valiant_2612174.cfm and https://amturing.acm.org/award_winners/lamport_1205376.cfm

So after suitable sleuthing, one can identify most of these computer-science Leslie's as individual persons. The ACM Digital Library provides a ready means for author identification, discerning such individuals as Leslie L. Miller and Leslie Jill Miller; readily handling hyphenated last names; and uniquely tracking one Leslie S. Smith (a retired computer science professor at University of Stirling). There were 37 male Leslie's including the two Turing awardees [graphed as p(F)=0.05] and a total of 83 female Leslie's [graphed as p(F)=0.95]. Thirteen Leslie's were not readily gender identified [so for them p(F)=0.5].[31] (To state the obvious, this data indicates publicly recognized or ascribed gender and is no reliable guide to an individual's gender identity.) Together, the Leslie's published 478 papers across 1970–2020. Figure 2 plots them all, by their authors' p(F), year-by-year, with the numbers of papers in a year indicated by the bubbles' areas. A reference line in green reminds us that NamSor computes Leslie's p(F) as 0.874, incorrectly presuming that Leslie's were always female.

Male Leslie's published 242 articles throughout this 50-year period, with exactly half on each side of 1995; for male Leslie's the median year is 1995. Female Leslie's 220 contributions were not so evenly divided. Through 1995, female Leslie's had published 23 articles, around 10% of their total for the duration; for female Leslie's the median year is 2008. Female Leslie's publishing in the early years include Leslie Chalmer's "Laboratory verification of patient identity" (AFIPS 1971) considering legal and professional responsibilities raised by computer-based laboratory report systems at the University of Pennsylvania Medical School; subsequently Chalmers became a noted figure in computer security, chair of the American Banking Association's security and risk management division, and quoted by *New York Times* reporter John Markoff.[32] Additional female Leslie's publishing in the 1980s include Leslie Maltz, longtime director of the computer center at Stevens Institute of Technology, applications analyst Leslie McCain at SUNY–Buffalo, veteran industrial designer Leslie Becker, statistician and experiment-design expert Leslie (Lisa) Moore at Los Alamos, and Leslie Jill Miller who drafted the influential reference model for OSI.[33] Leslie Ann Goldberg is a prolific computer science professor at Oxford with 39 papers,[34] while Leslie Pack Kaelbling is a prominent computer science professor at MIT and founder of the *Journal of Machine Learning Research* (2000–).[35]

[31] This may not be by accident, since some women engineers purposely avoid disclosing their gender, according to Shelley K. Erickson, "Women Ph.D. Students in Engineering and a Nuanced Terrain: Avoiding and Revealing Gender," *Review of Higher Education* 35 no. 3 (2012): 355-374 at doi.org/10.1353/rhe.2012.0019

[32] Samuel Raymond, Leslie Chalmers, and Walter Steuber, "Laboratory Verification of Patient Identity," In Proceedings of the May 18-20, 1971, Spring joint computer conference (AFIPS '71 (Spring)). Association for Computing Machinery, New York, NY, USA, 265–270. DOI: https://doi.org/10.1145/1478786.1478823 ; Leslie Chalmers, "User Identification, Access Control, and Audit Requirements," In Proceedings of the 1984 annual conference of the ACM on The fifth generation challenge (ACM '84). Association for Computing Machinery, New York, NY, USA, 247 at doi.org/10.1145/800171.809639 ; John Markoff, "Scientists Devise Math Tool To Break a Protective Code," *New York Times* (October 3, 1991) at www.nytimes.com/1991/10/03/us/scientists-devise-math-tool-to-break-a-protective-code.html (identifying Chalmers as female).

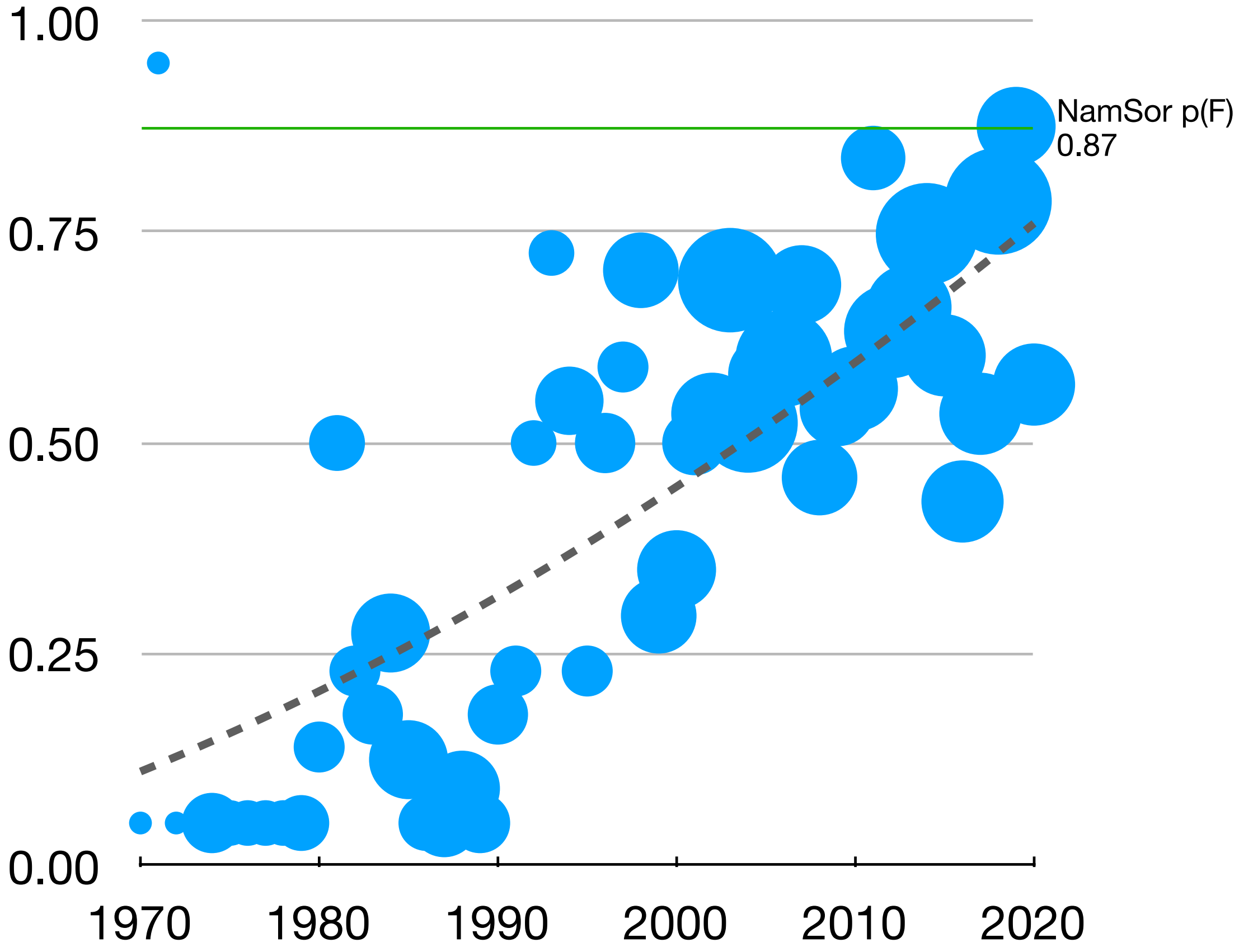

---

[33] Leslie Maltz, "Micros through mainframes: A successful implementation," In Proceedings of the 11th annual ACM SIGUCCS conference on User services (SIGUCCS '83). Association for Computing Machinery, New York, NY, USA, 19–26 at doi.org/10.1145/800041.801418 ; https://www.stevens.edu/news/need-speed-serial-entrepreneur-heads-firm-promises-more-powerful-computer ; and Leslie Jill Miller, "The ISO Reference Model of Open Systems Interconnection: A first tutorial," In Proceedings of the ACM '81 conference (ACM '81). Association for Computing Machinery, New York, NY, USA, 283–288 at doi.org/10.1145/800175.809901 ; Leslie J. McCain, "An audiovisual orientation to academic computing services," In *Proceedings of the 13th annual ACM SIGUCCS conference on User services: pulling it all together* (*SIGUCCS '85*). Association for Computing Machinery, New York, NY, USA, 21–34 at doi.org/10.1145/318741.318745 and www.acsu.buffalo.edu/~mccain/ (CV); https://dl.acm.org/profile/60000000162 and https://www.sparkawards.com/about/jury-alumni/leslie-becker/ (id); and https://dl.acm.org/profile/81100041828 and https://www.lanl.gov/search-capabilities/profiles/lisa-moore.shtml

[34] https://www.cs.ox.ac.uk/people/leslieann.goldberg/

[35] https://www.csail.mit.edu/person/leslie-kaelbling

The three computer-driven software tools assessed here firmly predict that all these Leslie's would be female (see table 1). The numerous male Leslie are therefore mis-labeled as female when using these gender-prediction tools. No software tool can reliably discern the male and female Leslies in this dataset; temporal awareness and contextual knowledge is needed. This is what human researchers excel at.

## Conclusions

Historians typically offer the methods of collective biography or prosopography to deal with larger samples of individuals. Arvid Nelsen used the method of prosopography in assessing a sample of 57 black computing professionals that he located in the pages of *Ebony* magazine.[36] His work has been cited in *Communications of the ACM* and in human-computer interaction research.[37] Often, such collective biographies can be used with moderate-sized populations (roughly 20 < N < 100).

Different methods are needed when populations expand (say N > 1,000). Some historical studies already combine traditional humanistic research methods with large-scale computer analyses. Nooney, Driscoll, and Allen assess early letters to a computer magazine (N = 1,285) acknowledging the use of the "open-source 'gender-guesser' Python package as a second heuristic for inferring the genders of *Softalk* letter writers . . . as a complement to human judgment."[38]

The consequences of the research presented here go in two directions. First, the impressive and published results from large population studies that use gender "prediction" software tools must be re-examined for their temporal appropriateness. So, too, results based on atemporal and/or non-contextual computing-software tools such as Gender-API, NamSor, and Genderize.io (see also forthcoming). Second, it seems likely that mixed methods combining humanistic research methods and contextual awareness with large-scale computational tools will be most valuable. Robust insights into the changing dynamics of gender bias, and other significant issues in computing's tangled relationship with politics, culture, and society, is at stake.

---

[36] R. Arvid Nelsen, "Race and Computing: The Problem of Sources, the Potential of Prosopography, and the Lesson of *Ebony* Magazine," *IEEE Annals of the History of Computing* 39 no. 1 (2017): 29-51 at doi: 10.1109/MAHC.2016.11.

[37] Fay Cobb Payton and Eleni Berki, "Countering the Negative Image of Women in Computing," *Communications of the ACM* 62, no. 5 (2019): 56-63 at doi.org/10.1145/3319422; and Alexandra To, Wenxia Sweeney, Jessica Hammer, and Geoff Kaufman, "'They Just Don't Get It': Towards Social Technologies for Coping with Interpersonal Racism," Proceedings ACM on Human-Computer Interaction 4, CSCW1, Article 24 (May 2020) at https://doi.org/10.1145/3392828

[38] Laine Nooney, Kevin Driscoll, and Kera Allen, "From Programming to Products: *Softalk* Magazine and the Rise of the Personal Computer User," *Information & Culture* 55 no. 2 (2020): 105-129, footnote 52, at muse.jhu.edu/article/757558.